\documentclass[twocolumn,english,groupedaddress, superscriptaddress,nofootinbib,aps,pre,10pt]{revtex4-1}
\usepackage[T1]{fontenc}
\usepackage[utf8]{inputenc}
\usepackage{gensymb}
\usepackage{textcomp}
\usepackage{amsmath,amsthm,mathtools,amssymb}
\usepackage{graphicx}
\usepackage{babel}
\usepackage{units}
\usepackage[normalem]{ulem}
\usepackage{tikz, pgfplots}
\pgfplotsset{compat=1.14}
\usepackage{siunitx}
\usepackage{filecontents}
\usepackage{color}
\definecolor{blue}{rgb}{0,0,1}

\definecolor{dgreen}{rgb}{0,0.5,0}

\definecolor{dred}{rgb}{0.5,0,0}

\definecolor{dyellow}{rgb}{0.75,0.75,0}

\definecolor{lightBlue}{rgb}{0.,0.5,0.5}

\definecolor{purple}{rgb}{0.5,0,0.5}

\renewcommand{\d}{\mathrm{d}}
\begin{document}

\title{Stochastic properties of the frequency dynamics in real and synthetic power grids}

\author{Mehrnaz Anvari}
\email{anvari@pks.mpg.de}
\affiliation{Max Planck Institute for the Physics of Complex Systems (MPIPKS), 01187 Dresden, Germany}

\author{Leonardo Rydin Gorj\~ao}
\email{l.rydin.gorjao@fz-juelich.de}
\affiliation{Forschungszentrum J\"ulich, Institute for Energy and Climate Research
- Systems Analysis and Technology Evaluation (IEK-STE), 52428 J\"ulich,
Germany}
\affiliation{Institute for Theoretical Physics, University of Cologne, 50937 K\"oln,
Germany}

\author{Marc Timme}
\email{marc.timme@tu-dresden.de}
\affiliation{Chair for Network Dynamics, Center for Advancing Electronics Dresden
(cfaed) and Institute for Theoretical Physics, Technical University
of Dresden, 01062 Dresden, Germany}
\affiliation{Network Dynamics, Max Planck Institute for Dynamics and Self-Organization (MPIDS), 37077 G\"ottingen, Germany}

\author{Dirk Witthaut}
\email{d.witthaut@fz-juelich.de}
\affiliation{Forschungszentrum J\"ulich, Institute for Energy and Climate Research
- Systems Analysis and Technology Evaluation (IEK-STE), 52428 J\"ulich,
Germany}
\affiliation{Institute for Theoretical Physics, University of Cologne, 50937 K\"oln,
Germany}

\author{Benjamin Sch\"afer}
\email{b.schaefer@qmul.ac.uk}
\affiliation{School of Mathematical Sciences, Queen Mary University of London, London E1 4NS, United Kingdom}
\affiliation{Chair for Network Dynamics, Center for Advancing Electronics Dresden
(cfaed) and Institute for Theoretical Physics, Technical University
of Dresden, 01062 Dresden, Germany}
\affiliation{Network Dynamics, Max Planck Institute for Dynamics and Self-Organization
(MPIDS), 37077 G\"ottingen, Germany}

\author{Holger Kantz}
\affiliation{Max--Planck Institute for the Physics of Complex Systems (MPIPKS), 01187 Dresden, Germany}

\begin{abstract}
The frequency constitutes a key state variable of electrical power grids. However, as the frequency is subject to several sources of fluctuations, ranging from renewable volatility to demand fluctuations and dispatch, it is strongly dynamic. Yet, the statistical and stochastic properties of the frequency fluctuation dynamics are far from fully understood.
Here, we analyse properties of  power grid frequency trajectories recorded from different synchronous regions. We highlight the non-Gaussian and still approximately Markovian nature of the frequency statistics. Further, we find that the frequency displays significant fluctuations exactly at the time intervals of regulation and trading, confirming the need of having a regulatory and market design that respects the technical and dynamical constraints in future highly renewable power grids.  
Finally, employing a  recently proposed synthetic model for the frequency dynamics, we combine our statistical and stochastic analysis and analyse in how far dynamically modelled frequency properties match the ones of real trajectories. 
\end{abstract}

\maketitle


\section{Introduction}
A stable electric power supply is essential for the functioning of our society \cite{Obama2013}. The ongoing energy transition towards renewable generation fundamentally changes the conditions for the operation of the power system \cite{Markard2018}. A better understanding of the dynamics, control, and variability of this highly complex system is needed to ensure stability in a rapidly changing environment \cite{Timme2015,Brummitt2013}.

The power grid frequency is the central observable for the control of AC electric power grids, as it directly reflects the balance of the grid:
A surplus of feed-in power increases the frequency and a shortage reduces the frequency \cite{Kundur1994}.
Observing the frequency of the power grid can thus provide deep insights into the dynamical stability of the grid as well as the operation of the control system and the economic dispatch of generators. In today's system strict operational boundaries are imposed on the frequency and the rate of change of frequency \cite{ENTSO-E2013}. For example, in the Central European power grid (CE), the stable operational boundary for frequency variations is set at $\pm 200$~\unit{Hz}. Moreover, if the frequency deviates more than $\Delta f = \pm 20$~\unit{Hz}, the existing control systems, i.e., primary and secondary control, are activated to compensate the imbalance in the power grid and to return the frequency to the nominal one \cite{Machowski2011}. 

These control mechanisms and operational boundaries are especially interesting when designing new grids involving concepts such as \emph{smart grids} \cite{Fang2012}, \emph{prosumers} \cite{kotler1986prosumer}, or \emph{microgrids} \cite{Lasseter2004}, and their interaction with the grid frequency.
Furthermore, due to the increased usage of renewable energies, synchronous machines are replaced by power electronics, such as inverters, posing additional challenges on ensuring frequency stability \cite{bottcher2019}.
Inverter-based generators do not have any innate inertia, leading to the frequency of the power grid becoming more volatile, unless additional stabilisers are included in the system  \cite{Milano2018}.

A more sophisticated analysis of the power-grid frequency dynamics is paramount, as all power generators and consumers have to ensure the stability of the grid in the presence of many effects simultaneously impinging on it. In such analyses it is both relevant to study existing power grids \cite{Weissbach2009} as well as to evaluate any forecasts and models of the frequency dynamics expected in future grids \cite{Gorjao2019}.

Despite the strict operational boundaries for frequency variations, numerous different sources of disturbances introduce measurable variations of the frequency over time. Important sources introducing fluctuations to the grid frequency include consumers, renewable energies, and the dispatch of power plants via the energy market. Recent research shows that today's demand fluctuations contribute substantially to uncertainties in the power balance \cite{Wood2013,ADRES-Concept,Tjaden2015}. 
Moreover, intermittent renewable energies influence the frequency firstly due to their stochastic and often non-Gaussian power feed-in \cite{Milan2013,Anvari2016}, and secondly due to the decreasing the inertia in the power grid, as mentioned above. Hence, to operate energy systems with a high share of renewable energies, a solid understanding of the impact of fluctuating feed-in on the grid's frequency is necessary.
Previous studies described the stochastic behaviour of the grid frequency using stochastic optimisation \cite{Zhao2013}, a simulated robustness analysis \cite{Anghel2007}, Fokker--Planck approaches \cite{Schaefer2017,Schaefer2017a}, or tracing the impact of wind feed-in on the grid frequency \cite{Haehne2018,wolff2019}.
However, the mathematical properties of the underlying stochastic process have not been studied comprehensively.

In addition to the aforementioned stochastic disturbances, trading affects the grid frequency by scheduled deterministic periodic events, e.g. dispatch actions on the energy market cause brief jumps of the frequency \cite{Weisbach2009,Schaefer2017a,Schafer2018b}. While deterministic disturbances have been observed for various grids \cite{Weisbach2009,Li2011}, no comprehensive model exists to describe the market interaction with the grid frequency quantitatively. We thus aim for a dynamical model of the power-grid frequency including the role of trading and regulator action in the power grid. Such model may help especially to plan future grids with a high share of renewable energies. Volatile renewable energies, such as wind and solar power, are unpredictable and thus cannot be used to balance the grid frequency following trading actions. Instead, it is fundamental to understand the interplay between the stochastic dynamics of unpredictable fluctuations and the deterministic characteristics of the energy market. 

Here, we first review essential statistical properties and the temporal evolution of the frequency of real-world power grids. Our approach provides a method to obtain bountiful information on the power-grid frequency that can be obtained from simple measurements. Next, we introduce our stochastic model to regenerate the frequency dynamics and explain how we estimate its parameters solely from the power-grid trajectory. Finally, we demonstrate how our model reproduces key aspects of the stochastic and deterministic behaviour of real trajectories.

\section{Power-grid frequency overview}\label{sec:II}
\begin{figure}
\includegraphics[width=1\columnwidth]{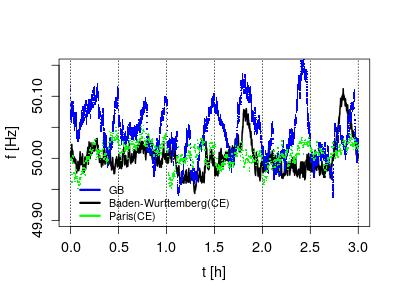}
\caption{The power grid frequency fluctuates over time, with differences between distinct regions. Displayed are three hours of frequency trajectories on March 1st for Paris, Baden-W\"urttemberg (both CE) and GB. The datasets are belong to 2015, 2016 and 2017, respectively, for Paris, GB and Baden-W\"urttemberg. Note that the Baden-W\"urttemberg and Paris data are from different years, while still displaying similar statistics.}
\label{SampleFreq}
\end{figure}

The power-grid frequency displays several characteristic features, such as non-Gaussian distributions, an exponential decay of the autocorrelation and regular impacts by trading \cite{Schaefer2017a}. We extend earlier studies by uncovering other stochastic properties of power-grid frequency, namely addressing the questions of Markovianity, linearity and stationarity of the data.
Specifically, we investigate the recorded frequency from Great Britain (GB) \cite{UK-Frequency2016}, and from two different regions in central Europe (CE). The two data samples of CE have been recorded in Paris (France) \cite{ReseauTransport2014-2019} and Baden-W\"urttemberg (South-West of Germany) \cite{Transnet}. The time resolution of data sets are $1$, $0.2$, and $1$ {\rm s}, respectively for GB, Paris, and Baden-W\"urttemberg. We analyse data spanning over one year: 2015 for France, 2016 for GB and 2017 for Baden-W\"urttemberg. The final section addresses the modelling following the data from Baden-W\"urttemberg. 
A direct observation of the frequency of the three samples (Great Britain, Paris, Baden-W\"urttemberg) during three arbitrarily chosen hours in March reveals substantial differences in the fluctuation patterns, see Fig.~\ref{SampleFreq}. The range of variations in GB is larger than in the other two frequency data sets. The reason being, the primary control in GB is only activated for frequency deviations of at least $\pm 200$~\unit{Hz}, while the other frequency sets belong to the CE grid, where control is activated at $\pm 20$~\unit{Hz}. Consequently the CE data set has smaller overall fluctuations and a lower standard deviation. 

\begin{figure*}
\centering
\includegraphics[width=1\columnwidth]{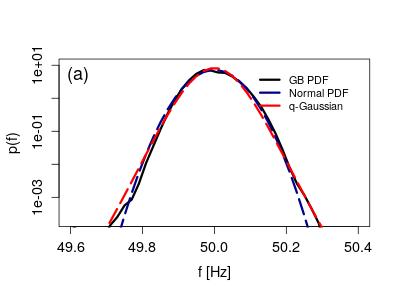}
\includegraphics[width=1\columnwidth]{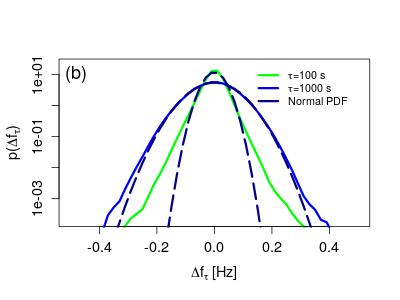}
\includegraphics[width=1\columnwidth]{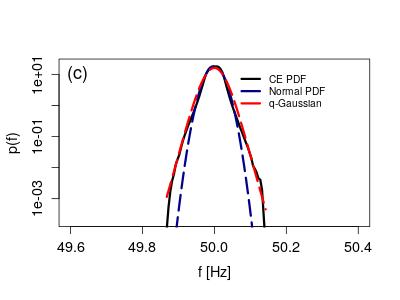}
\includegraphics[width=1\columnwidth]{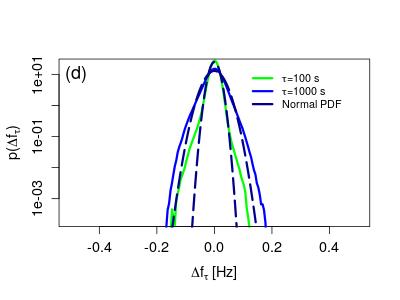}
\caption{Both the PDFs of the frequency and of the frequency increments display non-Gaussian features. We compare the PDF of the frequency with the most-likely Gaussian fit (blue curve) and q-Gaussian (red curve), for (a) the GB grid and (c) the CE grid evaluating the Baden-W\"urttemberg time series. We observe an asymmetry (non-zero skewness $\beta$) in the GB data with the deformation parameter $q=0.95$ and pronounced heavy tails (high kurtosis $\kappa$) in the CE data with $q=1.1$. Increment statistics in (b) GB and (d) CE grid were carried out for different time lags. Short-time lag ($\tau=100$)~{\rm s} displays more pronounced deviations from Gaussianity (dashed lines) than larger time lags.}
\label{PDF}
\end{figure*}

In contrast to many random processes, the values of the power grid frequencies  do not strictly follow Gaussian (normal) distributions \cite{Schaefer2017b, Kashima2015}. Instead, the distributions display heavy tails, where large deviations occur much more frequently than anticipated from a normal distribution.
In fig.~\ref{PDF}, the frequency and increment frequency distributions of GB and Baden-W\"urttemberg are shown. As both Paris and Baden-W\"urttemberg belong to the CE power grid, they have similar (but not identical) statistical properties. Therefore, for the rest of this section, we focus our analysis on the frequency measurements from Baden-W\"urttemberg as an example, and where we aim to refer to general statistic features, we refer to the CE grid. Comparing the frequency probability distribution function (PDF) with the best-fitting normal distribution, highlights the non-Gaussian properties of the frequency PDF of CE, which has a kurtosis $4.23$, fig.~\ref{PDF}(c). The kurtotsis, the normalised 4th moment, measures the heavy-tailedness of a distribution, see e.g. \cite{Samorodnitsky1994}. 
Any value  of the kurtosis larger than the that of a normal distribution ($\kappa_{normal}=3$) indicates heavy tails \cite{Westfall2014}. 
The frequency distribution for GB breaks the symmetry expected from a normal distribution and exhibits a skewness of $0.191$, see fig.~\ref{PDF}(a). The skewness, the normalised 3rd moment, $\beta$, measures how skewed, i.e., asymmetric, a distribution is. For a normal distribution, the skewness is zero. Furthermore, based on the shape of the PDFs, large deviations of the power grid frequency towards very low frequencies occur more often in the GB grid, while deviations towards higher frequencies are more common in the CE grid. We note that both skweness and kurtosis statistics depend on the sample size, but the observed non-Gaussian features are genuine since we do use large data sets with high sample frequency. Instead of normal distributions, the observed statistics is possibly better described by L{\'e}vy-stable or q-Gaussian distributions \cite{Schaefer2017a}.

The frequency increment statistics also display non-Gaussian features.
We estimate the probability to observe large fluctuations on short-time scales by computing frequency increments, i.e., $\Delta f_{\tau}=f(t+\tau)-f(t)$, see fig.~\ref{PDF}, panel (b) and (d), for $\tau=100$~{\rm s} and $\tau=1000$~{\rm s}, respectively. Next, we compare the observed increment probabilities with the best Gaussian fit: Frequency variations of the order of $210$~\unit{Hz} within $100$~\unit{Hz} occur in the GB frequency data set $10^{5}$ times more often than expected for Gaussian processes. For theBaden-W\"urttemberg data, frequency variations $\sim 60$~\unit{Hz} occur $100$ times more often compared to a Gaussian distribution. 
The increment frequency statistics indicates that the frequency on the short-time scale is particularly subject to large fluctuations. Potentially new control systems or market mechanisms are necessary to compensate the power imbalance in the power grid on short-time scales. In contrast, the shape of the frequency and frequency increment PDF become similar for larger time lags, such as $\tau=1000$, and the deviation from Gaussianity is not as extreme as for the short-time scale, see fig.~\ref{PDF}, panels (b) and (d). 

\begin{figure}
\centering
\includegraphics[width=1\columnwidth]{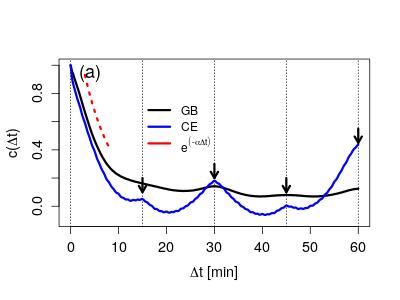}
\includegraphics[width=1\columnwidth]{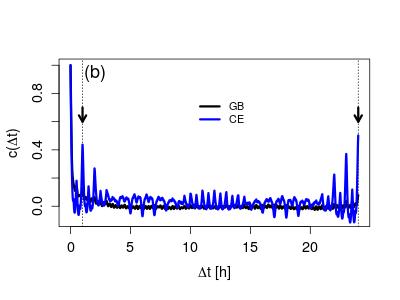}
\caption{Regular peaks in the auto--correlation demonstrate a mismatch between  power supply and demand (a) The autocorrelation $c(\Delta t)$ of GB and CE for a 1-hour lag period. The black arrows indicate the times of trading/dispatch actions after $15$ and $30$ minutes, which cause the peaks in the autocorrelation. The dotted red line reports the exponential decay of the autocorrelation in the first $10$ minutes. The inverse damping constants $\alpha^{-1}$ are estimated to be $\sim 385$ and $\sim 312$ {\rm s} for the GB and CE power grids, respectively. (b) The autocorrelation function $c(\Delta t)$ of the GB (black) and CE (blue) data sets for a 24-hours lag period. Regardless of regions, the initial exponential decay is followed by regular autocorrelation peaks. The black arrows highlight peaks of the autocorrelation after one hour and and also after $24$ hours, related to the periodicity of the frequency trajectory.}\label{ShortAutoCorr}
\end{figure}

To obtain more information from the frequency trajectory, we investigate the autocorrelation and its decay for the frequency data sets. The autocorrelation measures the correlation of a signal with itself at a later time. High correlation values indicate that a large signal is typically followed by still a large signal and vice versa. The power-grid frequency autocorrelation decays approximately exponentially as a function of the time lag $\Delta t$ for short-time lags, see \cite{Schaefer2017a} and fig.~\ref{ShortAutoCorr}. Several prototypical stochastic processes, such as the Ornstein--Uhlenbeck process, display a similar decay, following precisely an exponential function \cite{Gardiner1985}
\begin{equation}
    c(\Delta t)=\langle f(t)f(t+\tau)\rangle,
\end{equation}
\begin{equation}
    c^{OU}(\Delta t)=\exp(-\alpha \Delta t),
    \label{acf_OU}
\end{equation}
with a damping constant $\alpha$. While initially the system is highly correlated with its own history, this damping will cause a decorrelation. Naturally, distinct power grids will have their specific characteristic damping constant. A least squares fit of an exponential decay \eqref{acf_OU} to the data yields $\alpha^{-1}$ is $\sim 385$ {\rm s} for the GB grid and $\sim 312$ {\rm s} for the CE grid  respectively, see fig.~\ref{ShortAutoCorr}, panel (a).

Another feature of the autocorrelation are the regular peaks at every $15$ minutes, which are highlighted with black arrows in fig.~\ref{ShortAutoCorr}. These peaks are caused by a mismatch of power supply and demand  \cite{Weisbach2009,Schaefer2017b,Schafer2018b}.
In most electricity grids the operation of dispatchable power plants is scheduled in 1 hour blocks, where additional (shorter) $30$ minutes and $15$ minutes intervals might exist. Hence the generation curve is step--like, while the demand varies continuously. From step to step, the power balance rapidly switches from positive to negative or vice versa, leading to large deviations of the grid frequency, which become visible in the autocorrelation function, see also \cite{Gorjao2019}. In addition, daily routine, scheduled events, etc., contribute to an increased correlation every hour and 24 hours, see black arrows in fig.~\ref{ShortAutoCorr}, panel (b). Again, based on the specific regulations of different synchronous regions and their transmission system operators, the nature of the autocorrelation differs from region to region. For instance, the height of peaks in the GB autocorrelation in fig.~\ref{ShortAutoCorr}, panel (a), is visibly smaller than CE, which we attribute to a smaller trading and regulatory volume and overall larger stochastic fluctuation in GB. Consequently, the deterministic aspect of the frequency dynamics is diluted in GB. 

\begin{figure}
\centering
\includegraphics[width=1\columnwidth]{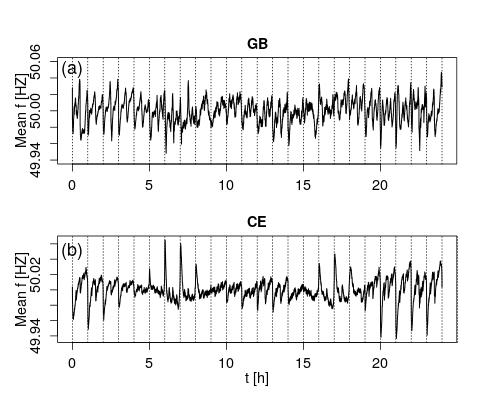}
\caption{Regular market activities induce periodic frequency jumps. Displayed is the frequency trajectory for (a) the GB grid, and (b) CE grid, averaged over all 366 days in 2016. We notice clear frequency jumps every hour, consistent with the previous observation of peaks in the autocorrelation function.}\label{Ave_Freq}
\end{figure}

Finally, to clearly demonstrate the impact of the energy trading market and related regulator actions on the frequency, we show the 
daily average frequency of both GB and CE in fig.~\ref{Ave_Freq}. 
The daily average frequency for every second is obtained by averaging over all days of the year.
The impact of the trading and regulation becomes  clear, as we observe sharp frequency jumps upwards or downwards every hour in both GB and CE. The direction of the jump and thereby the question whether the grid is displaying a shortage or a surplus of power is not random but also follows a deterministic pattern.

\begin{figure*}
\centering
\includegraphics[width=1\columnwidth]{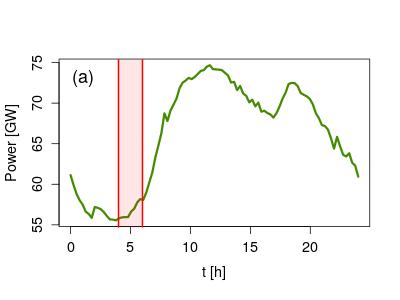}
\includegraphics[width=1\columnwidth]{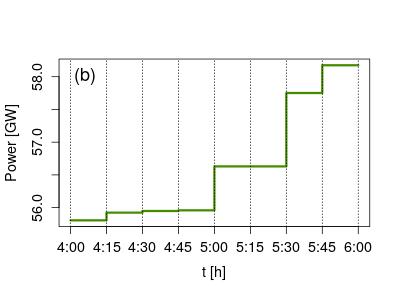}
\caption{
Discrete power dispatch leads to jumps of the scheduled power supply. 
(a) We display the real dispatch trajectory of the electricity supply in Germany in one day in 2017 \cite{ENTSOE_Demand}.
(b) The scheduled power jumps every $15$ minutes, as highlighted by the zoom on the two hours highlighted in red in (a).
Overall, the scheduled power supply approximates the changing demand throughout the day. Its discrete nature leads to jumps of the supply that has to be compensated by control mechanisms.
}
\label{fig:Generation-Demand-Illustration}
\end{figure*}

The market design is different for various synchronous grids or different countries within the same grid. For example, both the CE and the GB data display a periodicity of frequency jumps but the frequency dynamics within the CE grid appears more predictable. 
Frequency drops occur in the CE grid in each hour between $\text{20:00}$ and $\text{00:00}$, while the frequency clearly increases between $\text{06:00}$ to $\text{08:00}$ and $\text{16:00}$ to $\text{18:00}$. This pattern is linked to the slope of the demand curve. The step--like generation curve anticipates an increase or decrease of the demand \cite{Weisbach2009}. 
In case of rising demand, such as during the morning, an increasing amount of power is dispatched for each trading interval, see fig.~\ref{fig:Generation-Demand-Illustration}(b). Every 15 minutes the generation is increased to anticipate the demand by the consumers. These discrete changes in the supplied power  form the basis for the power mismatch in the synthetic frequency model discussed below.

\section{Stochastic properties}\label{sec:III}

Before we introduce a stochastic model for the power frequency dynamics, we perform some complementary tests to further characterise the underlying stochastic dynamics. Is the observed stochastic process stationary or nonstationary? Do we observe time symmetry, i.e., is the underlying process linear or nonlinear?
Does the process depend on its past or only on the current state, i.e., is the process Markovian?

\begin{figure}
\includegraphics[width=1\columnwidth]{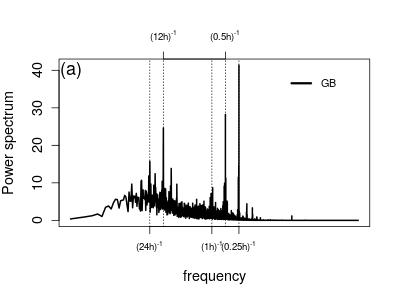}
\includegraphics[width=1\columnwidth]{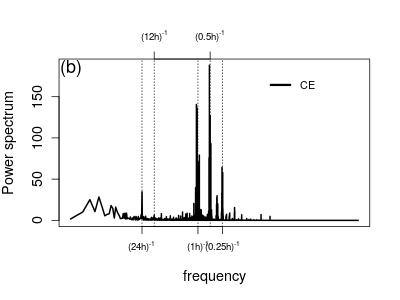}
\caption{Market activities and long time scales introduce non-stationarity. We plot the power spectrum of GB, panel (a), and of the CE data, panel (b). The spectra exhibit well-determined peaks before they decay on large-time scale. The dotted vertical lines show quarter-, half-, one, twelve and $24$ hour cycles (from right to left). 
}\label{Spectrum}
\end{figure}

{\bf{Stationary process:}} To test the reproducibility of the measured frequency we first investigate the stationarity for the data.
In the general definition, a probabilistic process is stationary if the probability of measured variables, in our case the probability of the frequency, does not depend on the time \cite{Kantz1997}. One of the standard methods to test the stationarity of a data set is analysing its spectrum. The sharp peaks in fig.~\ref{Spectrum} emphasise the existence of the periodicity on different time scales in the considered data. According to the spectrum, there are visible periods every quarter-, half-, one, twelve and $24$ hours in the grid frequency. This shows the nonstationary of the data on these time scales. However beyond $24$ hours, i.e., on longer time scales,
the spectrum is decreasing and consequently the data becomes stationary. 

There are other natural cycles influencing a power-grid system, such as the weekend-weekday pattern, as well as seasonal and yearly cycles. However, these cycles do not seem to leave a significant imprint in the spectrum of the power-grid frequency. Our 
stochastic model will focus on the intermediate time scale and hence include the characteristic daily dispatch and demand pattern, while neglecting longer-term processes.

\begin{figure*}
\centering
\includegraphics[width=1\columnwidth]{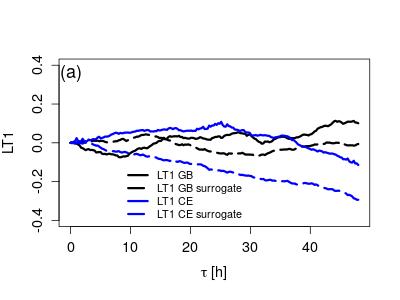}
\includegraphics[width=1\columnwidth]{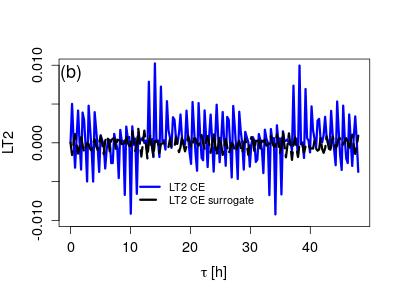}
\caption{The frequency trajectories display small non-linear effects. Panel (a) displays the $LT1$ results for the GB and CE frequency measurements. The dashed lines show the $LT1$ results for the surrogate datasets. The surrogate results act as a reference case of a linear model. Comparing the results of the original data with surrogate ones, we conclude both GB and CE are approximately linear. (b) $LT2$ results for the CE data. Surrogate (dashed black) and original data (solid blue) do differ more than when using $LT1$. This difference and the periodicity in $LT2$ are the signature of small nonlinear effects. 
\label{Linear-Nonlinear}}
\end{figure*}

{\bf{Linear process:}} Next, we investigate if there is any nonlinearity in the recorded power-grid frequency. For this purpose, consider the three-point autocorrelation of the frequency data as a measure of the time asymmetry in the data. If a time series is asymmetric in time, it is also nonlinear \cite{Kantz1997}. The following relations have been suggested to calculate the three-point autocorrelation for a data set \cite{Kantz1997}:  

\begin{equation}\label{time-Sym1}
    LT1 = \langle f(t)^2f(t+\tau)\rangle-\langle f(t)f(t+\tau)^2\rangle,
\end{equation}
\begin{equation}\label{time-Sym2}
   LT2= \langle(f(t)-f(t+\tau))^3\rangle/\langle(f(t)-f(t+\tau))^2\rangle,
   \end{equation}
where LT stands for linear test. A linear, and therefore time-symmetric, trajectory has both $LT1$ and $LT2$ sufficiently close to zero. Checking the validity of our results for a realistic process, we compare the original data to a surrogate time series, that provides a reference point of $LT1$ and $LT2$ for a linear process.
To generate the surrogate time series, we first take the Fourier transform (FT) of the original data and then randomise the phases. Finally, we employ an inverse FT to obtain the surrogate data.
With the described procedure we suppress any nonlinearity in the process, and therefore the surrogate data includes only the linear characteristics of the considered data \cite{Theiler1992}. 
The original data is linear if the $LT$ result of the original data lies within the value range of the $LT$ results of the ensemble of surrogate data. Here, instead of displaying the full ensemble of surrogate data in fig.~\ref{Linear-Nonlinear}, we have shown just an example for a surrogate data to avoid to obscure the figure.
Comparing the $LT1$ results of the surrogate datasets with the $LT1$ of the original datasets displays that the qualitative behaviour of both are equivalent, entailing that the processes approximately follow linear characteristics, for both the GB and the CE datasets, as seen in fig.~\ref{Linear-Nonlinear}, panel (a). Looking more closely at the $LT1$ for the CE surrogate data, which only includes the linear characteristics and fluctuations, we note that its deviation from zero are larger than $LT1$ for original CE data.  Investigating the value of $LT2$ for GB also confirms the linear characteristic of the data set. As the $LT1$ and $LT2$ results for GB are the same, we only show the $LT1$ results. However, for the CE data set, $LT2$ indicates that the data might not be strictly linear but displays small nonlinearities, as seen fig.~\ref{Linear-Nonlinear}, panel (b). 
As shown in fig.~\ref{ShortAutoCorr}, the effect of the market activity in CE is more regular and more severe than in GB, therefore we suspect that the nonlinearity in CE data is caused by the regular jumps in the frequency trajectory. When devising our model, we will therefore approximate the weakly non-linear process as linear.

\begin{figure*}
\centering
\vspace{-8em}
\includegraphics[width=1.\columnwidth]{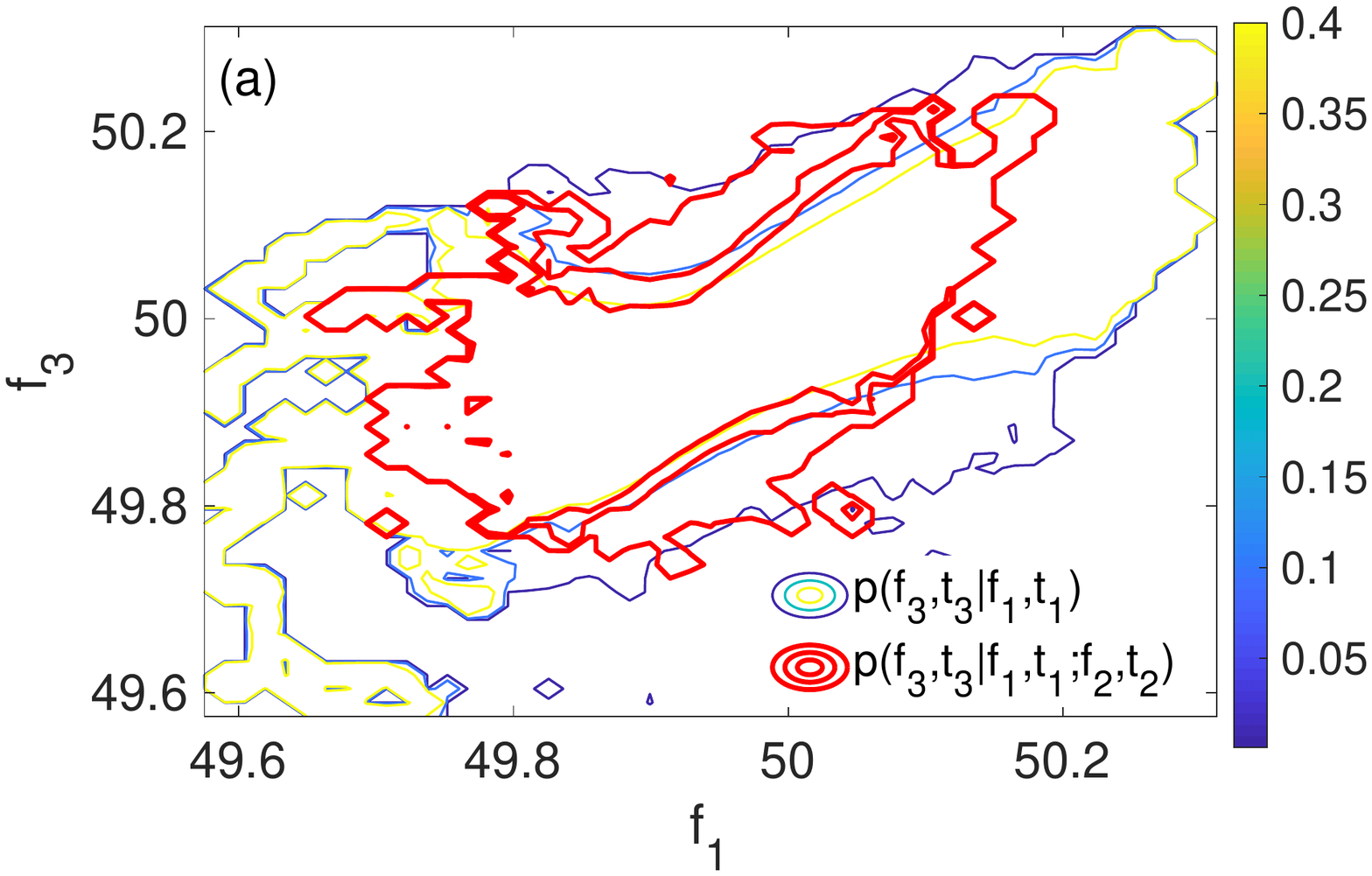}
\includegraphics[width=1.\columnwidth]{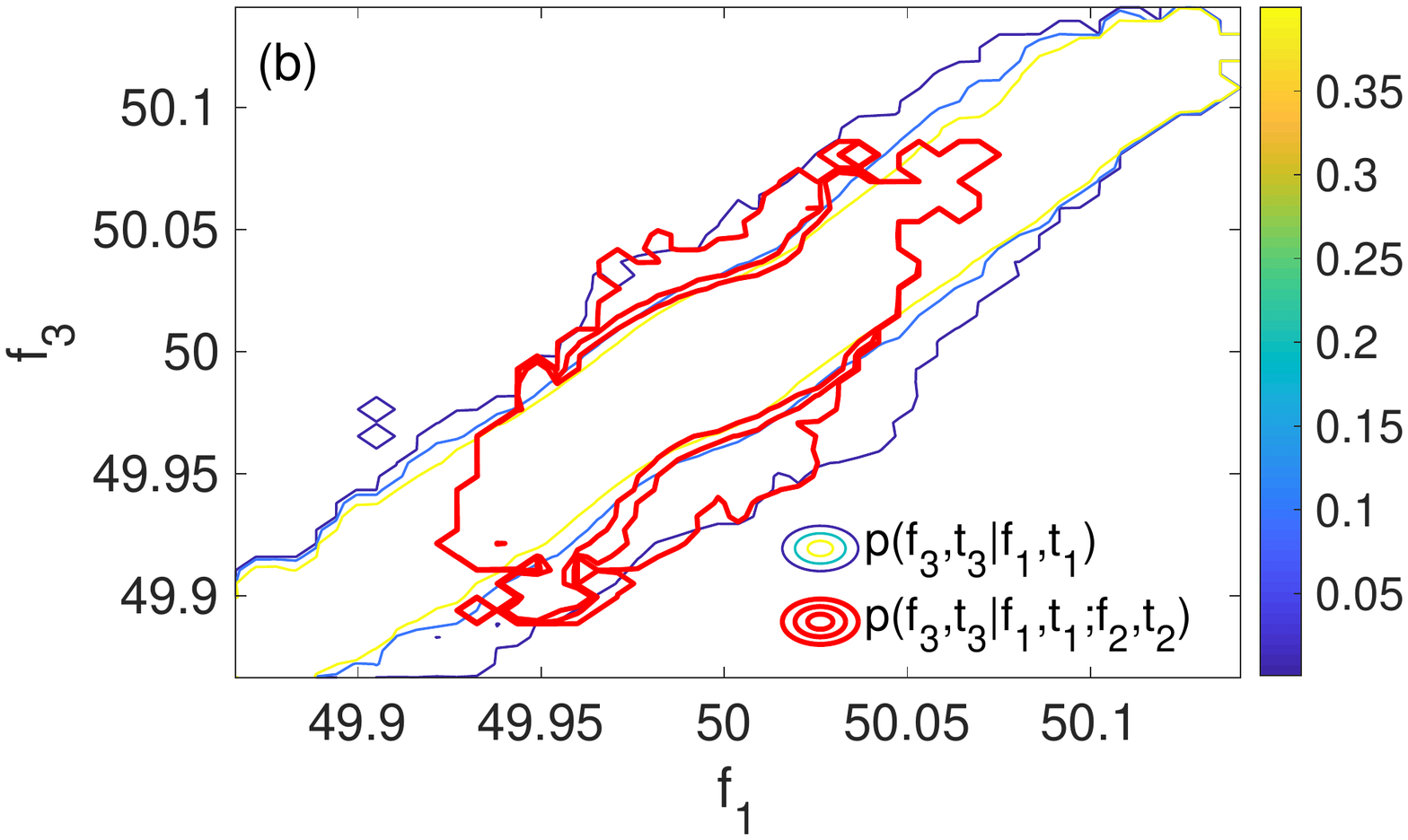}
\vspace{-8em}
\caption{The Markovian nature of the real data is confirmed by a Chapmann--Kolmogorov test for (a) the GB grid and (b) the CE grid using the Baden-W\"urttemberg dataset. The proximity of the contour lines of $p(f_3,t_3|f_1,t_1,f_2,t_2)$ (red contour) and $p(f_3,t_3|f_1,t_1)$ (coloured contour) show the validity of Chapman--Kolmogorov test for the frequency datasets. The time $t_1$ is chosen to contain $10$ data points to show the contours clearly. Next, the times $t_2$ and $t_3$ are multiples of $t_1$, chosen as $2t_1$ and $3t_1$,  respectively. \label{fig:Markovian_Property}}
\end{figure*}

{\bf{Chapmann--Kolmogorov test:}} A fundamental property of stochastic processes is whether future states only depend on the current state or whether they have memory. In other words, whether the process is Markovian or not. A well-known approach to evaluate whether a process is Markovian is the Chapmann--Kolmogorov test \cite{Gardiner1985}. According to the Chapmann--Kolmogorov test, the conditional PDFs of Markovian processes obey the following equation
\begin{equation}
p(f_3,t_3|f_1,t_1)=\int{p(f_3,t_3|f_2,t_2)p(f_2,t_2|f_1,t_1)df_2},
\label{Chap-Kol}
\end{equation}
where $t_3>t_2>t_1$. 
To test the Markovianity for the data, instead of employing directly eq.~\eqref{Chap-Kol}, one considers its $2D$ and $3D$ conditional PDF. As shown in fig.~\ref{fig:Markovian_Property}, $p(f_3,t_3|f_1,t_1)$ and $p(f_3,t_3|f_2,t_2;f_1,t_1)$ match approximately, implying the power grid frequency is mostly Markovian. Any stochastic model for the power frequency should therefore be Markovian as well.

\section{Stochastic model}
We now introduce a synthetic model for the power-grid frequency as a stochastic, mostly linear, and Markovian process.
The stochastic model presented here aims at reproducing essential features of a power grid, as well as its statistical characteristics, and consists of three independent systems: Firstly, the intrinsic deterministic dynamics of the power grid, including primary and secondary control. Secondly, it embodies as well a stochastic signal or noise, as evidenced by the aforementioned frequency trajectories \cite{Schafer2018b}.
Thirdly, we model the sudden power imbalance arising after the dispatch actions by implementing an appropriate deterministic function: We make use of historic dispatch data and apply it using a a step function of the power. Other functions, such as artificial steps or saw-tooth like functions are also possible.

\begin{figure*}[t]
\centering
\includegraphics[width=1\columnwidth]{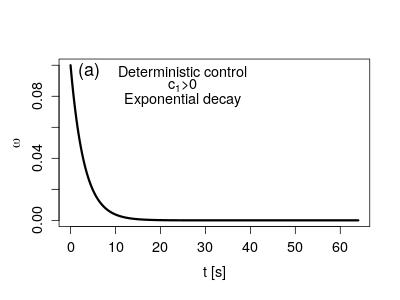}
\includegraphics[width=1\columnwidth]{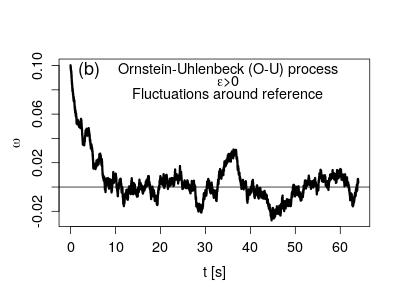}
\includegraphics[width=1\columnwidth]{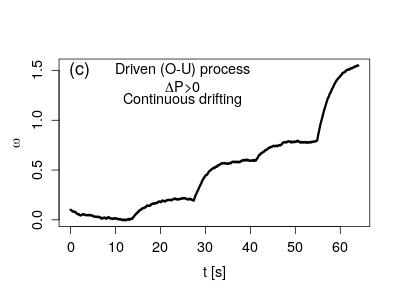}
\includegraphics[width=1\columnwidth]{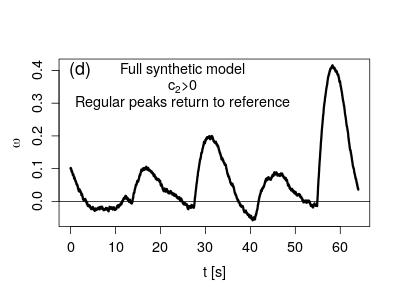}
\caption{All terms of the synthetic model \eqref{freq_derivation} are necessary to reproduce the frequency trajectory. We plot the angular velocity $\omega$ as a function of time when using the synthetic frequency model \eqref{freq_derivation} but setting individual parameters to $0$. Parameters are chosen for pure illustrative purpose and we set $\omega (0)=0.1$ as initial condition.
(a): Including only primary control leads to a pure exponential decay of the angular velocity. (b): Adding non-zero noise $\epsilon$, we recover an Ornstein--Uhlenbeck process. (c): including a step function for the power imbalance $\Delta P$ leads to a continuously drifting Ornstein--Uhlenbeck process. (d): Finally, including secondary control guarantees that the angular velocity returns back to the reference. Parameters are: $\epsilon =0.001/s^2$, $c_1=0.005/s$, $c_2=0.00003/s^2$, $\Delta P=0.004/s^2$ at every hour and half or a quarter of it every $30$ or $15$ and $45$ minutes respectively.}
\label{fig:ModelOverview}
\end{figure*}

Instead of the actual frequency, we use the bulk angular velocity relative to the reference frequency of $50$~\unit{Hz}, $\omega=2\pi(f-50~\unit{Hz})$ to express our model. 
Contrary to network analysis on power grids \cite{Filatrella2008,Rohden2012}, we have only access to frequency measurements on the global scale and therefore average over all nodes to obtain the averaged (bulk) frequency and angular velocity \cite{Ulbig2014} $\omega=\frac{1}{M}\sum_{i=1}^NM_i\omega_i$, where $M=\sum_{i=1}^NM_i$ is the total inertia of all nodes and $N$ is the number of nodes in the power grid.
Typically, the frequency at each node is very close to the bulk frequency throughout the grid, with fluctuations indicating the gross power balance.
Notable exceptions are high-frequency disturbances, which are typically localised \cite{zhang2018fluctuation,haehne2019propagation}, or inter-area oscillations, where energy is oscillating from one part of the grid to another one.
The synthetic model of the frequency dynamics is discussed in detail in \cite{Gorjao2019}.
It is given as a linear stochastic differential equation:
\begin{equation}\label{freq_derivation}
\begin{aligned}
\frac{\d\omega}{\d t} =  -c_1 \omega  - c_2 \theta  +  \Delta P_{\text{ext}} + \epsilon\xi,
\end{aligned}
\end{equation}
with bulk angle $\theta$ and its derivative $\d {\theta}/ \d t =\omega$. Furthermore, $\Delta P_{\text{ext}}$ is the exogenous influence on the power balance, i.e., the trading or dispatch impact of the power imbalance, $\epsilon$ and $\xi$ are the noise amplitude and Gaussian white noise function, respectively.
Finally, $c_1$ is the magnitude of the fast-acting primary control, while $c_2$ is the magnitude of the secondary control which acts slower and lasts longer than primary control.
We illustrate the contribution of the different terms of the synthetic model~\eqref{freq_derivation} in fig.~\ref{fig:ModelOverview}.

The full model is displayed in fig.~\ref{fig:ModelOverview}(d): In case of an abundance of generation, i.e., a sudden positive $\Delta P_\text{ext}$, the frequency increases above the reference ($50$~\unit{Hz}).
The primary control $c_1$ mitigates the sudden rise of the frequency and quickly stabilises the frequency, but not at the nominal value of $50$~\unit{Hz}.
Subsequently, the secondary control slowly restores the frequency back to its reference of $50$~\unit{Hz}. According to the time schedule of control systems, we assume that the primary control acts faster than secondary control, and consequently $c_1\gg c_2$ \cite{Kundur2004,tchuisseu2018curing}. 

Furthermore, the nature of the dispatch structure $\Delta P_{\text{ext}}$ must be specified. The generation of each power plant (the dispatch) is rapidly adapted by the operators, e.g. based on trading at the European Energy Exchange. 
As discussed in detail at the end of sec.~\ref{sec:II}, the operation of dispatchable power plants is scheduled at fixed intervals. As we have shown in fig.~\ref{fig:Generation-Demand-Illustration} the power generation can increase or decrease every $15$ minutes, which we model approximately as a step function, with potentially different step sizes at the full hour, $30$ minutes, or $15$ minutes intervals. 
On the other hand, data of power generation in different regions or countries is generally available, and can be implemented directly in the model.
In the model presented here, we extracted the power generation in Germany for the equivalent month of December 2017, and used this as the power balance $\Delta P_{\text{ext}}$ \cite{ENTSOE_Demand}.

Before we compare results of the synthetic model with the real data, we need to determine suitable parameters.
Details are given in \cite{Gorjao2019} on how to estimate the parameters from a given frequency trajectory.
In short, the noise amplitude $\epsilon$ is estimated based on the stochastic fluctuations around the observed frequency trajectory, while the power imbalance $\Delta P_{\text{ext}}$ is directly read from the rise or sag of the frequency at the scheduled time points of dispatch, which are proportional to the missing or exceeding amount of power.
(Notice that in our case we include the real power generation from Germany for December 2017, thus circumvent extracting the power generation $\Delta P_{\text{ext}}$ from the data).
Primary control $c_1$ is recovered by studying the process' affinity to revert its trajectory to the dispatched power and secondary control $c_2$ is estimated from the frequency recovery rate to the nominal value after a scheduled action \cite{Gorjao2019}.

\section{Quantitative comparison between model and data}
To evaluate the stochastic model described above, we generated one month of synthetic data with a one second resolution, mirroring the CE data from December, 2017.
The parameters for the synthetic model \cite{Gorjao2019} are estimated from the $1$-second resolution data-series provided by \cite{Transnet} and their values are shown in table~\ref{Parameter}.
The data for the power generation for the month of December, 2017, in Germany can be found in \cite{ENTSOE_Demand}. 

\begin{table}
\begin{center}
\caption{The parameters for the synthetic model for CE, December, 2017}\label{Table2}\vspace{0.05cm}
\begin{tabular}{ccc}
\hline\hline~\vspace*{-0.9em}\\[0.1em] 
~\quad~ $\epsilon$ ($s^{-2}$) ~\quad~ & ~\quad~  $c_1$ ($s^{-1}$) ~\quad~ &  ~\quad~ $c_2$ ($s^{-2}$) ~\quad~ \\
  \hline~\vspace*{-0.9em}\\[0.1em] 
   $0.00107$ & $0.00915$ & $0.00003$ \\ 
\hline
\end{tabular}\label{Parameter}
\end{center}
\end{table}

\begin{figure*}
    \centering
    \includegraphics[width=0.45\textwidth]{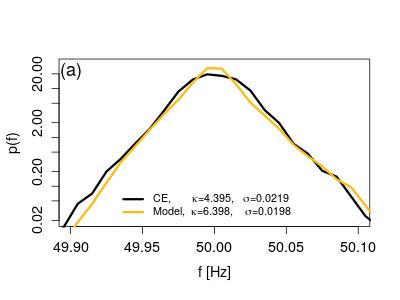}
    \includegraphics[width=0.45\textwidth]{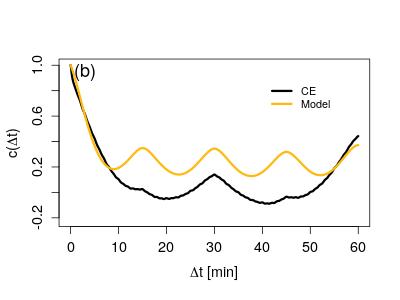}
    \caption{
    The synthetic model captures important features of the real data, including trading peaks and heavy tails.
    (a): The probability distributions of the frequency data from CE in 2017 (black), compared to our synthetic model (yellow). Both display distinct heavy tails with kurtosis $\kappa>3$.
    (b): The autocorrelation function of the frequency initially decays and then displays regular peaks at the trading intervals.}
    \label{fig:Histo_Correlation_Leo}
\end{figure*}

Now, we repeat most of our statistical and stochastic analyses to compare how well the synthetic model reproduces the original data.
First, we note that the general shape of the PDF (see fig.~\ref{fig:Histo_Correlation_Leo}(a)) and autocorrelation (see fig.~\ref{fig:Histo_Correlation_Leo}(b)) do agree well between the model (yellow) and the empirical data (black).
Both the model and the data display heavy tails, i.e., the aforementioned deviation from Gaussianity.
Furthermore, the autocorrelation function of the synthetic model captures the regular peaks, due to the changing dispatch.
The decay of the autocorrelation function is approximately described by the current model.
Both results emphasise the enormous impact of the energy market activity and dispatch structure on the dynamics and stability of the power system.

\begin{figure}
    \centering
    \vspace{-7em}
    \includegraphics[width=1\columnwidth]{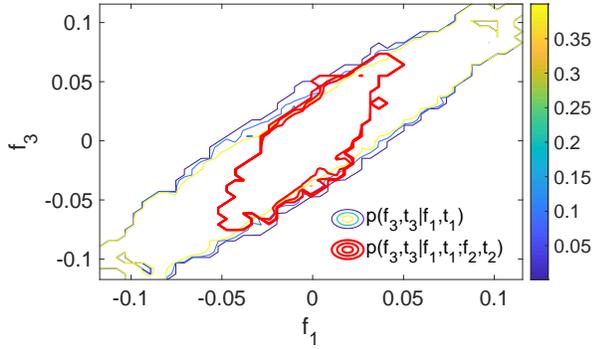}
    \vspace{-10em}
    \caption{Chapmann--Kolmogorov test confirms the Markovian nature of the synthetic model.
    The test used one month of synthetic CE data generated by \eqref{freq_derivation}.
    }
    \label{fig:Chapman_Kolmogorov_Synthetic}
\end{figure}

\begin{figure}
    \centering
    \includegraphics[width=1\columnwidth]{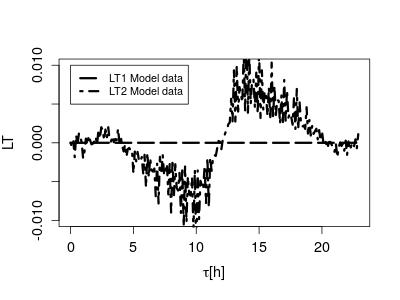}
    \caption{
    The synthetic model is approximately linear. We apply the linear  tests on the time series generated by the synthetic model: $LT1$ shows linear characteristics for CE dataset, however $LT2$ reports a small nonlinearity also found in the real data of the CE power-grid frequency.}
    \label{fig:Linear_Test_Synthetic}
\end{figure}

Consistent with our modelling assumptions, we find that the synthetic model is Markovian, based on a Chapman--Kolmogorov test, see fig.~\ref{fig:Chapman_Kolmogorov_Synthetic}.
Similarly, we do observe that both $LT1$ and $LT2$ results show that the synthetic model has compatible characteristics with the real one, i.e., while the $LT1$ reports a linear process, $LT2$ results show a small nonlinearity in the synthetic, c.f. fig.~\ref{fig:Linear_Test_Synthetic}.
As we discussed in sec.~\ref{sec:III}, this nonlinear behaviour is likely linked to the regular trading in the CE power grid.

We again emphasise that our model addresses the dynamics on the intermediate time-scale of the frequency, i.e., approximately $30$ seconds to a few hours.
On shorter time-scales, our model neglects: (i) dynamical behaviour of rotating machines, (ii) non-trivial stochastic noise, (iii) network dynamics, and (iv) momentary reserve vs. primary control.
Moreover, the switching in trading is not instantaneous as we have assumed in the 
Similarly, our model does not include all effects acting on larger time-scales, for example: 
(i) feed-in of wind and solar power, which determines how much inertia exists in the system and how much the generation side fluctuates.
(ii) dispatch of power plants determined on the spot market, such as the European Energy Exchange (EEX). This is especially relevant for areas where no historic market data is available or forecasts are attempted.
In order to capture these effects, we would need a full fledged market model plus meteorological input for the weather data.

\begin{figure}
    \centering
    \includegraphics[width=1\columnwidth]{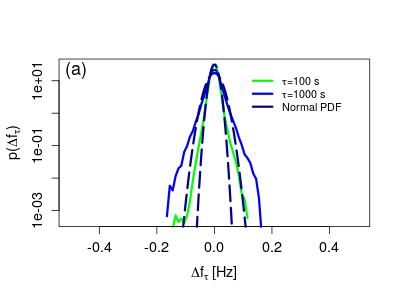}
    \includegraphics[width=1\columnwidth]{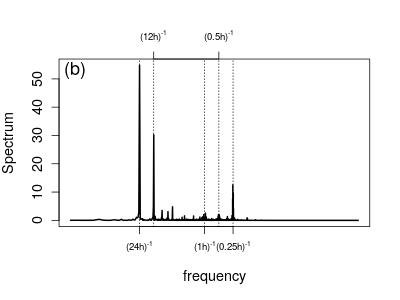}
    \caption{
    Increment and spectral analysis of the synthetic model are consistent with the real data. (a) The increment statistics of the synthetic data shows non-Gaussian characteristics similar to the real one.
    (b) The spectrum of the synthetic frequency trajectory reports large peaks at the trading times, while it decays to zero for longer time scales.
    The dotted vertical lines show respectively, quarter-, half-, one, twelve, and $24$ hours from right to left.}
   \label{fig:Spectrum_Synthetic}
\end{figure}

The spectral analysis and the increment statistics of the synthetic data are shown in fig.~\ref{fig:Spectrum_Synthetic}.  Similarly to fig.~\ref{PDF}, in fig.~\ref{fig:Spectrum_Synthetic}(a) the frequency increment statistics of the generated data also display non-Gaussian features on short-time scales as the real data. 
The spectrum of the synthetic frequency trajectory displays several pronounced peaks, which are mostly consistent with the trading times of the model, i.e., quarter-hourly, twelve hours and twenty four hours (c.f. fig.~\ref{fig:Spectrum_Synthetic}).

\section{Discussion}
In summary, we have presented an analysis of the statistics of power-grid frequency dynamics, with an emphasis on non-standard behaviour.
In particular, we have shown the non-Gaussian nature of the power-grid frequency fluctuations in the aggregated and increments statistics, which includes heavy tails.
Furthermore, we have demonstrated that the power-grid frequency trajectory is adequately described as a Markovian process and that it shows small nonlinear effects.
Regulatory and trading events introduce some periodicities in the system. 
Finally, based on the observed properties, we have constructed a synthetic model that captures not only the aggregated statistics in terms of the histogram but also qualitatively reproduces the observed autocorrelation decay, correlation peaks due to market activity, increment statistics, and spectral properties of the real data \cite{Gorjao2019}.
The model is well suited to understand the energy-market effects on power-grid frequency on intermediate time scales and goes beyond previous studies focusing on a description \cite{Weisbach2009,Schafer2018b} of trading or a stochastic theory \cite{Schaefer2017a}.
We here focused on a statistical and stochastic analysis of real-world frequency dynamics, with a comparison to the presented model.
The analysis of the synthetic model is consistent with our modelling assumptions, in that it is approximately Markovian and displays small nonlinear and periodic market effects.
Furthermore, we found that the observed heavy tails of the frequency distributions arise mainly due to trading actions, impacting not only the frequency temporally close to the market action but also several minutes later. This is clear since we only applied Gaussian noise to an otherwise linear dynamics. Only the deterministic trading actions can therefore cause the non-Gaussian properties.
The spectral and increment properties of the synthetic model also approximate the original real-world data, which confirms again the effect of the trading market on the frequency dynamics.
It is worth to re-iterate that the presented model is conceptually simple, easy to implement, and includes a minimum set of adjustable parameters. Therefore, we explicitly did not model the machine dynamics, noise on very short time scales or a detailed market and dispatch model.
Some alternative model approaches, involving more fitting parameters are explored in \cite{Gorjao2019}.

Concluding, our analysis of power-grid frequency dynamics and the stochastic model we presented, including a structured comparison, may help
to better understand the interplay of the internal dynamics and external disturbances of electric-power systems and to develop improved simulation models.
A thorough understanding of this interplay is a prerequisite for the design and optimisation of future electricity markets, as well as regulatory and control schemes.
For instance, the current market design in the continental European grid regularly causes substantial frequency deviations when the dispatch is adjusted every $15$ minutes such that primary control has to be activated on a regular basis. A smoother change of the dispatch could reduce these frequency deviations and reduce stress onto the primary and secondary control system \cite{Weissbach2009}. Alternatively, frequency regulations could be adapted in a way that the typical frequency deviations due to the changing dispatch are tolerated while exceptional cases are identified and handled by the control system.
Our structured analyses (Markov, stationary and linearity properties) and model may offer a powerful and versatile framework to study these questions, in particular because the model, while still simple, simultaneously captures essential features of the interplay of internal dynamics, control, and market activity.
The presented analysis and modelling framework can thus contribute to the design of future power system, reducing the necessity for control actions and saving costs.

The model can further be used to assess the frequency stability of future power-grid structures, including in particular microgrids \cite{Fang2012} or low-inertia grids \cite{Milano2018}.
Traditional dynamical stability analyses focus on local and global stability of fixed points and the impact of large isolated disturbances such as the sudden shutdown of power plant.
In comparison, the impact of ongoing stochastic disturbances on grid stability has received less attention.
As evidenced in this study, the regulatory system and market design may have play an important role for these external stochastic effects.

We kept the model as simple as possible to reproduce key features of the frequency time series such as the histogram and the autocorrelation. Future research could naturally extend the model to better match the spectrum or long-time autocorrelation. 
Furthermore, one could investigate particular intervals of the power grid trajectory, e.g. high- vs. low-demand intervals, such as weekdays vs. weekends. Additional stochastic investigations could further quantify the agreement between real data and the synthetic model, e.g. by investigating higher-order N-point statistics, going beyond our current 2-point statistics (increments).



\begin{acknowledgments}
We gratefully acknowledge support from the Federal Ministry of Education and Research (BMBF grant no. 03SF0472, 03EK3055), the Helmholtz Association (via the joint initiative ``Energy System 2050 - A Contribution of the Research Field Energy'' and the grant no. VH-NG-1025) and the German Science Foundation (DFG) by a grant toward the Cluster of Excellence ``Center for Advancing Electronics Dresden'' (cfaed). This project has received funding from the European Union’s Horizon 2020 research and innovation programme under the Marie Sk\l{}odowska-Curie grant agreement No 840825.

\end{acknowledgments}


\bibliographystyle{naturemag}
\bibliography{References}

\end{document}